\documentclass[12pt]{article}
\usepackage{amsmath,amssymb}
\begin{document}

\begin{titlepage}

\setlength{\baselineskip}{18pt}

                            \begin{center}
                            \vspace*{2cm}
        \large\bf  Phase space measure concentration for an ideal gas\\

                            \vfill

              \small\sf ANTHONY J. CREACO$^{\ast}$, \ \ \ \ NIKOS KALOGEROPOULOS$^{\dagger}$\\

                            \vspace{0.2cm}

 \small\sf  Department of Science\\
            BMCC-The City University of New York\\
            199 Chambers St., New York, NY 10007, USA\\
                            \end{center}

                            \vfill

                     \centerline{\normalsize\bf Abstract}
                            \vspace{1mm}

\normalsize\rm\noindent\setlength{\baselineskip}{18pt} We point
out that a special case of an ideal gas exhibits concentration of
the volume of its phase space, which is a sphere, around its
equator in the thermodynamic limit. The rate of approach to the
thermodynamic limit is determined. Our argument relies on the
spherical isoperimetric inequality of L\'{e}vy and Gromov.\\

                             \vfill

 \noindent\sf PACS: \ \ \ \ \ \  02.30.Cj, \ \ 02.40.Ky, \ \ 05.20.Gg\\
\noindent\sf Keywords: \ Ideal gas, Measure concentration, Isoperimetric inequalities.\\

                             \vfill

\noindent\rule{11.5cm}{0.2mm}\\
\begin{tabular}{ll}
\noindent\small\rm E-mail: & \small\rm $^\ast$ acreaco@bmcc.cuny.edu\\
                  & \small\rm $\dagger$ nkalogeropoulos@bmcc.cuny.edu, nkaloger@yahoo.com
\end{tabular}

\end{titlepage}


                                 \newpage

\setlength{\baselineskip}{18pt}

                    \centerline{\sc 1. \  Introduction}

                               \vspace{3mm}

The concentration of measure phenomenon [1] is an important
concept in contemporary analysis and measure theory. It was very
effectively used by V.D. Milman in his proof of Dvoretzky's
theorem [2] on the existence of almost spherical sections of
convex bodies and by many others ever since [1]. Roughly speaking,
the concentration of measure states as the dimension of the
underlying geometric structure, be it a subset of Euclidean or a
Banach space, a manifold, a metric space etc, approaches infinity,
most of the measure considered in such a structure can be found
concentrated in a subset of it. Such a subset would be considered
``small" by the casual observer. Dually, measure concentration can
be seen as a property of functions of a large number of variables
with small oscillations which turn out to be almost constant. This
interpretation goes back to P. L\'{e}vy [3]. The geometric concept
of ``observable diameter" and related quantities introduced in
[4] is also motivated by this observation.\\

Isoperimetric inequalities are some of the most studied geometric
inequalities [5], especially during the last century and a half.
Roughly speaking, isoperimetric inequalities  provide bounds to
the relation between the volume of a domain and the area of its
boundary. A physical motivation is to prove the observation that
the shape of a droplet, or bubble, in ordinary flat space is
spherical and to explore what this optimal shape would be in other
situations such as Riemannian spaces of constant or variable
curvature, metric-measure spaces etc. Despite the best efforts of
decades, such minimizers are known only in the case of the
Euclidean space \ $\mathbb{R}^n$ [6], \ the sphere \ $S^n$ [7] \
and the hyperbolic space \ $H^n$ [8] \ equipped with their usual
metrics of constant sectional curvature \ $0, +1$ \ and \ $-1$ \
respectively. The concentration of measure phenomenon is
controlled through inequalities, whose geometric incarnation has a
strong isoperimetric flavor. The present work use the fact that
spherical caps are the isoperimetric domains/minmizers in \ $S^n$ [7].\\

In this letter we present an application of the measure
concentration on \ $S^n$ \ and the related isoperimetric
inequality of P. L\'{e}vy [3] which was generalized much later by
M. Gromov [7], to a particular case of an ideal gas. To be more
specific, we consider an ideal non-relativistic gas which is
placed inside a one-dimensional thermally isolated box, i.e.
inside a line segment. We assume that initially this gas contains
\ $n-1$ \ (point) particles/molecules. We then add one extra
particle/molecule which is stationary in the frame of reference of
the box. The total energy of the system remains the same before
and after we add this $n$-th particle. We continue this process,
ad infinitum, and we try to understand the behavior of the volume
of the phase space of the system in such a limit. The phase space
has a compact part which is a sphere. We find that as \ $n$ \
increases,  the volume of such a sphere concentrates around its
equator  The result that we find is that, in the thermodynamic
limit \ $n\rightarrow\infty$, \ the micro-canonical partition
function of the gas receives its dominant contribution from a
tubular neighborhood of the equator of width proportional to \
$n^{-\frac{1}{2}}$. \ Thus, we provide a geometric description of
the system and, through it, an expression for the rate at which
the system approaches equilibrium. This letter should be
considered as a counterpart of [11] where a similar type of
arguments was used to determine the rate of approach to
equilibrium of a parity-violating system. \\

In Section 2, we start the mathematical description of such a
system by calculating the relative volume of a tubular
neighborhood of the equator of a sphere and compare it to the
volume of the sphere itself. We find  that, to a first order
approximation, that this ratio is zero, as is intuitively
expected. In Section 3 we show how the situation changes
dramatically when a more careful, second order, approximation to
such a ratio is made. Here the concentration of measure becomes
manifest. In Section 4 we discuss how Sections 2 and 3 can be
applied to the specific case of the ideal gas that we are
considering. Section 5 contains some further comments and some
conclusions that can be drawn from such an analysis.\\


                              \vspace{3mm}

\centerline{\sc 2. \ Relative volumes of spheres; first order
approximation}

                               \vspace{3mm}

Consider a system of $n+1$ non-relativistic particles on a line.
The phase space of such a system is \
$\mathbb{R}^{n+1}\times\mathbb{R}^{n+1}$. \ If the total energy \
$E$ \ of the system is conserved, and assume that, for any reason,
the location of the individual particles does not affect the value
of \ $E$, \ then the phase space reduces to  \
$S^{n}\times\mathbb{R}^{n+1}$. \ We focus our attention on the
compact part of the phase space, the \ $n$-sphere \ $S^n$, \
because, eventually, the contribution of the potential function,
which depends on the generalized coordinates only, will turn out
to be a constant in the micro-canonical partition function. Such a
constant will not be relevant to our arguments, so we will
eventually disregard it. We assume, for simplicity, that \ $S^n$ \
is equipped with its round metric \ $\rho$ \ induced from the
Euclidean metric of \ $\mathbb{R}^{n+1}$. \ It is well-known that
the geodesic segment joining $x_1, x_2 \in S^n$ is an arc of a
great circle of \ $S^n$. \ Let the corresponding convex angle
subtended by such a geodesic arc at the center of the sphere be \
$\psi$. \ Then
\begin{equation}
     \rho(x_1, x_2) = \psi
\end{equation}
The Euclidean distance in the ambient space \ $\mathbb{R}^{n+1}$ \
is expressed in terms of \ $\psi$ \ as
\begin{equation}
     \| x_1-x_2 \| \ = \ 2 \sin \frac{\psi}{2}
\end{equation}
and   the relation between \ $\rho(x_1,x_2)$ \ and \ $\| x_1-x_2
\|$ \ is
\begin{equation}
    \frac{2}{\pi} \ \rho (x_1, x_2)  \ \leq \ \| x_1-x_2 \| \ \leq \ \rho
    (x_1, x_2)
\end{equation}
Let \ $d\mu$ \ denote the infinitesimal Riemannian measure on \
$S^n$ \ induced from the Lebesgue measure of \ $\mathbb{R}^{n+1}$.
\ We have assumed that \ $S^n$ \ is trivially embedded in
$\mathbb{R}^{n+1}$, which is parametrized by \ $x^i, \
i=1,\ldots,n+1,\ \ x^i\in \mathbb{R}$ \ and $S^n$ is parametrized
by the standard spherical coordinates \ $\theta_i, \ i=1,\ldots,
n$ \ with \ $ 0\leq\theta_j<\pi, \ j=1,\ldots, n-1, \ \
0\leq\theta_n<2\pi$ \ namely
\begin{eqnarray}
 x^1 & = & \sin\theta_1\sin\theta_2 \cdots
 \sin\theta_{n-1}\sin\theta_n \nonumber \\
 x^2 & = & \sin\theta_1\sin\theta_2 \cdots
 \sin\theta_{n-1}\cos\theta_n \nonumber \\
     & \vdots  &                                \\
 x^n & = & \sin\theta_1\cos\theta_2 \nonumber \\
 x^{n+1} & = & \cos\theta_1 \nonumber
\end{eqnarray}
The Jacobian of (4) gives
\begin{equation}
 d\mu = d\theta_n \prod_{j=1}^{n-1}\sin^{n-j}\theta_j \ d\theta_j
\end{equation}
Let \ $U_{\epsilon/2}(S^{n-1})$ \ denote an
$\epsilon/2$-neighborhood of the equator \ $S^{n-1}\subset S^n$ \
namely
\begin{equation}
 U_{\epsilon/2}(S^{n-1}) = \{ x\in S^n : \rho(x,S^{n-1})
< \frac{\epsilon}{2} \}
\end{equation}
and let \ $vol(U_{\epsilon/2}(S^{n-1}))$ \ be its volume obtained
from \ $d\mu$. \ Then
\begin{equation}
vol(U_{\epsilon/2}(S^{n-1})) = 2\pi
\int_{\frac{\pi-\epsilon}{2}}^{\frac{\pi+\epsilon}{2}}
\sin^{n-1}\theta_1 \ d\theta_1 \
\prod_{j=2}^{n-1}\int_0^{\pi}\sin^{n-j}\theta_j \ d\theta_j
\end{equation}
which gives
\begin{equation}
vol(U_{\epsilon/2}(S^{n-1})) \ = \ vol(S^{n-1}) \
\int_{\frac{\pi-\epsilon}{2}}^{\frac{\pi+\epsilon}{2}}
\sin^{n-1}\theta_1 \ d\theta_1
\end{equation}
Setting \ $\phi = \theta_1 - \frac{\pi}{2}$, \ and using the fact
that \ $\cos\phi$ \ is an even function on \ $\mathbb{R}$, \ we
find
\begin{equation}
vol(U_{\epsilon/2}(S^{n-1})) \ = \ vol(S^{n-1}) \
2\int_0^\frac{\epsilon}{2} \cos^{n-1}\phi \ d\phi
\end{equation}
We are interested in determining the ratio
\begin{equation}
 \frac{vol(U_{\epsilon/2}(S^{n-1}))}{vol(S^n)} \ = \
 \frac{vol(S^{n-1})}{vol(S^n)} \
      2\int_0^\frac{\epsilon}{2} \cos^{n-1}\phi \ d\phi
\end{equation}
To complete the calculation we have to determine \ $
2\int_0^\frac{\epsilon}{2} \cos^{n-1}\phi \ d\phi.$ \ We can
evaluate it recursively or we can prove by induction, a
posteriori, that [12] for \ $m\in\mathbb{N}$,
\begin{equation}
  \int_0^{\frac{\epsilon}{2}} \cos^{2m+1}\phi \ d\phi \ = \
        \frac{1}{2^{2m}}\sum_{k=0}^m {2m+1 \choose k}
        \frac{\sin [(2m-2k+1)\frac{\epsilon}{2}]}{2m-2k+1}
\end{equation}
and
\begin{equation}
  \int_0^{\frac{\epsilon}{2}} \cos^{2m}\phi \ d\phi \ = \
   \frac{1}{2^{2m}}{2m \choose m}\frac{\epsilon}{2} +
   \frac{1}{2^{2m-1}}\sum_{k=0}^{m-1}{2m \choose k}
     \frac{\sin [(2m-2k)\frac{\epsilon}{2}]}{2m-2k}
\end{equation}
In the sequel, we examine the case \ $n=2m+1$. \ The case \ $n=2m$
\ can be analyzed in the same way so we omit it, for brevity. We
see that
\begin{equation}
 2 \int_0^{\frac{\epsilon}{2}} \cos^{n-1}\phi \ d\phi =
 2\int_0^{\frac{\epsilon}{2}} \cos^{2m}\phi \ d\phi =
 \frac{1}{4^m}\frac{(2m)!}{(m!)^2}\epsilon + \frac{1}{4^{m-1}}
 \sum_{k=0}^{m-1}\frac{(2m)!}{k!(2m-k)!}\frac{\sin[(2m-2k)\frac{\epsilon}{2}]}{2m-2k}
\end{equation}
We are eventually interested in a small \
$\frac{\epsilon}{2}$-neighborhood of the equator, with \
$\epsilon$ \ chosen to satisfy
\begin{equation}
|(m-k)\epsilon|\ll 1, \ \ \ \mathrm{as} \  \ m\rightarrow\infty
\end{equation}
so a Taylor series expansion of the sine function gives, to first
order in \ $(m-k)\epsilon$
\begin{equation}
 2 \int_0^{\frac{\epsilon}{2}} \cos^{2m}\phi \ d\phi \ = \
  \frac{1}{4^m}\frac{(2m)!}{(m!)^2}\ \ \epsilon + \frac{1}{4^{m-1}}
 \sum_{k=0}^{m-1}\frac{(2m)!}{k!(2m-k)!}\frac{(2m-2k)\frac{\epsilon}{2}}{2m-2k}
\end{equation}
We then use the well-known observation that as \ $m$ \ increases \
$m!$ \ increases much faster, so the maximum of \ ${2m \choose k}$
\ is attained when \ $k=m$, \ which results in
\begin{equation}
  2 \int_0^{\frac{\epsilon}{2}} \cos^{2m}\phi \ d\phi \ \leq \
   \frac{1}{4^m}\frac{(2m)!}{(m!)^2} \ \epsilon +
   \frac{2}{4^m}(m-1)\frac{(2m)!}{(m!)^2} \ \epsilon
\end{equation}
which eventually gives
\begin{equation}
  2 \int_0^{\frac{\epsilon}{2}} \cos^{2m}\phi \ d\phi \ \leq \
   \frac{2m-1}{4^m} \ \frac{(2m)!}{(m!)^2} \ \epsilon
\end{equation}
We express the factorial through the gamma function \ $\Gamma(x)$
\ as [12] \ $\Gamma(m+1) = m!$ \ and can rewrite (17) as
\begin{equation}
  2 \int_0^{\frac{\epsilon}{2}} \cos^{2m}\phi \ d\phi \ \leq \
    \frac{2m-1}{4^m} \ \frac{\Gamma(2m+1)}{[\Gamma(m+1)]^2} \
    \epsilon
\end{equation}
Using
\begin{equation}
\Gamma(x+1)\ = \ x\Gamma(x), \ \ \ \ \ x\in\mathbb{C}\setminus
\{-1, -2, \ldots \}
\end{equation}
we find
\begin{equation}
 2 \int_0^{\frac{\epsilon}{2}} \cos^{2m}\phi \ d\phi \ \leq \
  \frac{4m-2}{m4^m} \ \frac{\Gamma(2m)}{[\Gamma(m)]^2} \ \epsilon
\end{equation}
Substituting \ $m=(n-1)/2$, \ we get
\begin{equation}
  2 \int_0^{\frac{\epsilon}{2}} \cos^{n-1}\phi \ d\phi \ \leq \
  \frac{n-2}{(n-1)2^{n-3}} \ \epsilon \
         \frac{\Gamma(n-1)}{[\Gamma(\frac{n-1}{2})]^2}
\end{equation}
It is well-known that
\begin{equation}
 vol(S^{n-1}) = \frac{2\pi^{\frac{n}{2}}}{\Gamma(\frac{n}{2})}
\end{equation}
which gives
\begin{equation}
  \frac{vol(S^{n-1})}{vol(S^n)} \ = \
  \frac{\Gamma(\frac{n+1}{2})}{\Gamma(\frac{n}{2}) \Gamma(\frac{1}{2})}
\end{equation}
Combining (10) and (23), we find
\begin{equation}
  \frac{vol(U_{\epsilon/2}(S^{n-1}))}{vol(S^n)} \ \leq \
  \frac{n-2}{(n-1)2^{n-3}} \ \epsilon \
  \frac{\Gamma(\frac{n+1}{2})\Gamma(n-1)}{\Gamma(\frac{1}{2})
  \Gamma(\frac{n}{2}) [\Gamma(\frac{n-1}{2})]^2}
\end{equation}
Using once more (19) and that \ $\Gamma(\frac{1}{2})= \sqrt{\pi}$,
\ (24) reduces to
\begin{equation}
  \frac{vol(U_{\epsilon/2}(S^{n-1}))}{vol(S^n)} \ \leq \
   \frac{n-2}{2^{n-2}} \frac{\epsilon}{\sqrt{\pi}}
   \frac{\Gamma(n-1)}{\Gamma(\frac{n}{2})\Gamma(\frac{n-1}{2})}
\end{equation}
Substituting the Legendre duplication formula [12]
\begin{equation}
  \Gamma(2x) \ = \ \frac{2^{2x-1}}{\sqrt{\pi}} \ \Gamma(x) \ \Gamma
  (x+\frac{1}{2})
\end{equation}
in (25), we find
\begin{equation}
 \frac{vol(U_{\epsilon/2}(S^{n-1}))}{vol(S^n)} \ \leq \
 \frac{(n-2)\epsilon}{\pi}
\end{equation}
We have already assumed in (14) that, essentially \
$|n\epsilon|\ll 1$. \ In this approximation, we find
\begin{equation}
 \lim_{n\rightarrow\infty} \frac{vol(U_{\epsilon/2}(S^{n-1}))}{vol(S^n)} \ =
 \ 0
\end{equation}
This is a result that someone intuitively expects to hold. The
volume of a small tubular neighborhood of the equator is much much
smaller than the volume of the whole sphere, even as \ $n$ \
increases without an upper bound. Revisiting this derivation we
see at least two points which are not quite satisfactory. First,
we were ``too generous" in our attempts to find an upper bound for
\ $2 \int_0^{\frac{\epsilon}{2}} \cos^{2m}\phi \ d\phi$ \
appearing in (10). We replaced \ $\sin[(2m-2k)\epsilon/2]$ \ with
its first order Taylor series approximation  \ $(2m-2k)\epsilon/2$
\ in (15). \ Second, we uncritically substituted all combinatorial
terms \ ${2m \choose k}$ \ with their maximum \ ${2m \choose m}$ \
in (16). The result is a very crude upper bound in (27), which is
not sufficiently sensitive to distinguish small deviations from
zero of \ $\frac{vol(U_{\epsilon/2}(S^{n-1}))}{vol(S^n)}$, \
in case they exist.\\

                               \vspace{5mm}


\centerline{\sc 3. \ Relative volumes of spheres; second order
approximation}

                               \vspace{3mm}

The state of affairs changes dramatically by sharpening the upper
bound of (13). This is done by working at a level of accuracy
equivalent to a second order approximation of \
$\sin(m-k)\epsilon$, \ which would be naively expected to vanish.
The approach of the last few paragraphs, which relies on the
explicit calculation of the volumes that we want to compare from
the outset, cannot be straightforwardly extended to help us reach
our goal. This is due to the fact that there is no obvious better
upper bound either for \ $\sin (m-k)\epsilon$ \ or for \ ${2m
\choose k}$ \ that can sharpen the inequality (16) which we need
to use. A second-order approximation in its argument exists only
for a cosine, but not for a sine function. The logic that we
follow is:  we first find a second-order upper bound for the
cosine function appearing in (10), and only then we perform the
required integrations.\\

One works as follows [9]. As in (6), let \ $B(x,\epsilon) = \ \{
y\in S^n : \rho(y,x)<\epsilon \}$. \ This represents
 a  ``cap" of radius \ $\epsilon$ \ on \ $S^n$. \  Instead of
the relative volume of the ``equatorial ring"  \ $\frac{vol
(U_{\epsilon/2}(S^{n-1}))}{vol(S^n)}$ \ that we considered in
(10), let us consider now
\begin{equation}
  \frac{vol(B(x,\frac{\pi}{2}+ \epsilon ))}{vol (S^n)}
\end{equation}
This represents the percentage  of the total volume that is
occupied by spherical ``cap" encompassing the equator. We  try to
find a bound for
\begin{equation}
\alpha(\epsilon, n) = 1 - \frac{vol(B(x,\frac{\pi}{2}+ \epsilon
))}{vol (S^n)}
\end{equation}
This function expresses the ``percentage" of the volume of \ $S^n$
\ which is left out after subtracting the volume of the spherical
``cap" defined above from \ $S^n$. \ According to the general
volume formulae (7),(8),(22), \ $\alpha(\epsilon, n)$ \ can be
expressed as
\begin{equation}
 \alpha(\epsilon, n) = \frac{\int_{\frac{\pi}{2}+\epsilon}^{\frac{\pi}{2}}
     \sin^n\theta \
    d\theta}{\int_0^{\pi}\sin^n\theta \ d\theta}
\end{equation}
Substituting \ $\phi = \theta - \frac{\pi}{2}$, \ as in (9) and
setting \ $I_n = \int_0^{\pi}\cos^n\phi \ d\phi$, \ (31) can be
rewritten as
\begin{equation}
\alpha(\epsilon, n) \ = \ \frac{\int_{\epsilon}^{\frac{\pi}{2}}
     \sin^n\phi \
    d\phi}{\int_0^{\pi}\sin^n\phi \ d\phi}
\end{equation}
Changing the variable of integration to \ $w = \phi \sqrt{n}$, \
we get
\begin{equation}
 \alpha(\epsilon, n) \ = \
 \frac{1}{2I_n\sqrt{n}}\int_{\epsilon\sqrt{n}}^{\frac{\pi}{2}\sqrt{n}}
    \cos^n(\frac{w}{\sqrt{n}}) \ dw
\end{equation}
Comparing the Taylor series expansions of \ $\cos w$ \ and of \
$\exp (-w^2/2)$ \ in \ $[0, \frac{\pi}{2}]$, \ we see that \ $\cos
w \leq \exp (-w^2/2)$, \ so (33) gives
\begin{equation}
  \alpha(\epsilon, n) \ \leq \frac{1}{2I_n\sqrt{n}}
    \int^{(\frac{\pi}{2}-\epsilon)\sqrt{n}}_0
    e^{-\frac{w^2}{2}} \ dw
\end{equation}
By setting \ $t = w+\epsilon\sqrt{n}$ \ we find that
\begin{equation}
  \alpha(\epsilon, n) \ \leq \frac{1}{2I_n\sqrt{n}}
    \int_{\epsilon\sqrt{n}}^{\frac{\pi}{2}\sqrt{n}}
    e^{-\frac{(t+\epsilon\sqrt{n})^2}{2}} \ dt
\end{equation}
which implies
\begin{equation}
  \alpha(\epsilon, n) \ \leq \ \frac{e^{-\frac{\epsilon^2n}{2}}}{2I_n\sqrt{n}}
    \int_0^{\infty} e^{-\frac{t^2}{2}} \ dt
\end{equation}
which using the fact that \ $\int_0^{\infty} \exp(-\frac{t^2}{2})
\ dt = \sqrt{\frac{\pi}{2}}$ \ gives
\begin{equation}
   \alpha(\epsilon, n) \ \leq \
   \frac{1}{I_n\sqrt{n}} \ \sqrt{\frac{\pi}{8}} \ \
   \exp(-\frac{\epsilon^2n}{2})
\end{equation}
The straightforwardly derived recursion relation \ $(n+2)I_{n+2} =
(n+1)I_n$ \ implies that
\begin{equation}
  \sqrt{n+2} \ I_{n+2} \ = \ \frac{n+1}{\sqrt{n+2}} \ I_n \geq
  \sqrt{n} I_n
\end{equation}
Obviously \ $I_1 \geq 1$ \ and \ $\sqrt{2}I_2\geq 1$, \ therefore
$\sqrt{n}I_n \geq 1, \forall \ \ n\in\mathbb{N}$, \ which gives
\begin{equation}
\frac{vol(B(x,\frac{\pi}{2}+ \epsilon ))}{vol (S^n)}  \ \geq \
  1-\sqrt{\frac{\pi}{8}} \ \exp (-\frac{\epsilon^2n}{2})
\end{equation}
Let \ $U\subset S^n$ \ with \ $vol U/vol(S^n) \geq 1/2$ \ and \
$\epsilon >0$. \ For any $\epsilon$-neighborhood \ $U_{\epsilon}$
\ of \ $U$, \ the classical L\'{e}vy-Gromov isoperimetric
inequality for \ $S^n$ \ [3] - [7] states that
\begin{equation}
 \frac{vol(U_{\epsilon})}{vol(S^n)} \ \geq \
    \frac{vol(B(x,\frac{\pi}{2}+ \epsilon ))}{vol (S^n)}
\end{equation}
Combining the above statements we arrive at the concentration
inequality
\begin{equation}
  \frac{vol(U_{\epsilon}(S^{n-1}))}{vol(S^n)} \ \geq \
    1-\sqrt{\frac{\pi}{8}} \ \exp (-\frac{\epsilon^2n}{2})
\end{equation}
Its implication is immediate, although counter-intuitive: as the
dimension \ $n$ \ of the unit sphere \ $S^n$ \ becomes very large,
all its volume is concentrating in a thin ring of width \
$\epsilon \sim \frac{1}{\sqrt{n}}$ \ around its equator. An
intuitive objection to this conclusion is the following; \ $S^n$ \
has an infinite number of great circles \ $S^{n-1}$ \ at the \
$\epsilon$-neighborhood in each of which the volume is
concentrated, according to (41). Let's pick one such great circle
with its corresponding \ $\epsilon$-neighborhood. Each \
$\epsilon$-neighborhood of every other great circle should have
volume zero, a fact that appears to contradict (41) which does not
discriminate in favor of one among the great circles and their \
$\epsilon$-neighborhoods. This inequality also seems to violate
the sphere being a symmetric space, consequently its points being
connected through the transitive action of its isometry group.
Since the volume is a Hausdorff measure directly derived from the
round metric \ $\rho$ \ of \ $S^n$, \ it appears
that is should also possess the symmetries of \ $\rho$. \\

This apparent contradiction can be partly resolved, if we consider
what happens to the volume of \ $S^n$ \ for large \ $n$. \
Substituting the Stirling approximation
\begin{equation}
   \Gamma(x+1) \ \sim \ \sqrt{2\pi x} \ \left(\frac{x}{e}\right)^x, \ \ \
   \mathrm{for} \ \  x\rightarrow\infty
\end{equation}
in (22) and using that
\begin{equation}
   \lim_{x\rightarrow\infty} \left(1+\frac{k}{x}\right)^x \ = \
   e^k, \ \ \forall \ k\in\mathbb{R}
\end{equation}
we find
\begin{equation}
   vol(S^{n-1}) \ \leq \ \sqrt{2}{\pi}
    \left(\frac{2e\pi}{n^{1-\frac{1}{n}}}\right)^\frac{n}{2}
\end{equation}
which implies that
\begin{equation}
     \lim_{n\rightarrow\infty} vol S^{n-1} \ = \ 0
\end{equation}
as fast as \ $n^{-\frac{n}{2}}$. \ A volume approaching zero means
that the $n$-dimensional manifold in question, \ $S^n$ \ in our
case, ``collapses" to one of lower dimension, from a
measure-theoretical viewpoint. This is odd, considering that the
metric properties of \ $S^n$  \ remain those of an $n$-dimensional
space. By comparing (41) and (45), we conclude that the volume of
\ $S^n$ \ decreases much faster toward zero than the relative
volume \ $\frac{vol(U_{\epsilon}(S^{n-1}))}{vol(S^n)}$ \ so,
measure-theoretically, \ $S^n$ \ is effectively reduced to \
$S^{n-1}$. \ Finally, it is probably worth mentioning, for
completeness, that this volume decline of \ $S^n$ \ toward zero as
a function of \ $n$ \ does not contradict the topology of \ $S^n$:
\ indeed, one can  argue [13] that the infinite sphere \
$S^{\infty}$ \ is contractible, hence topologically trivial. The
discrepancy between our intuition and the results of this section
should make us be very careful when trying to extend already
understood concepts of a fixed dimension, to the case of such a
dimension approaching infinity. Indeed, as we just saw, in the
latter case, which is a particular example in the asymptotic
theory of normed spaces, our low-dimensional geometric intuition
may misguide us if we try to make predictions or support arguments
by relying on it too much. This is not totally surprising as
geometry, proper, actually refers to spaces of fixed (Hausdorff)
dimension. By contrast, we are dealing with calculations of
non-trivial limits of sequences of inequalities in  spaces where
the dimension is a free parameter, and not totally surprisingly
new phenomena occur in such limits.\\

 It may also be worth noticing that in (41) the relative volume
of the tubular neighborhood of the equator declines in a Gaussian
way as a function of \ $\epsilon$. \ Because of this, (41) can  be
seen as providing a geometric model for the law of large numbers
[4].\\


                              \vspace{5mm}

\centerline{\sc 4. \ Application to an ideal gas}

                               \vspace{3mm}

Consider a gas, in one spatial dimension for simplicity, of \
$n-1$ \ identical particles of unit mass, with momenta \ $p_i, \
i=1,\ldots,n-1$ \ and interacting with each other through the
potential \ $\Phi(x_1,\ldots,x_{n-1})$. \ The Hamiltonian of such
a system is given by
\begin{equation}
     \mathcal{H} \ = \ \sum_{i=1}^{n-1} \frac{p_i^2}{2} \ + \ \Phi(x_1,
      \ldots, x_{n-1})
\end{equation}
We put this gas inside a thermally isolated one-dimensional box (a
line segment) of volume (length) \ $V$, \ whose walls (endpoints)
are perfectly reflecting. The equilibrium statistical behavior of
such a system is determined by calculating the micro-canonical
distribution
\begin{equation}
          \tau(p_i, x_j): \mathrm{const}, \ \ \ \ 0\leq i,j \leq
          n-1
\end{equation}
on the constant energy \ $E$ \ hyper-surface of phase space with
the normalization
\begin{equation}
          \int_{\mathcal{H}=E} \tau (p_i, x_j) = 1
\end{equation}
 For an ideal gas
\begin{equation}
 \Phi (x_1, \ldots, x_{n-1})=0
\end{equation}
Then, the phase space of the system is \
$V^{n-1}\times\mathbb{R}^{n-1}$. \ We have used the fact that the
total energy \ $E$ \ of the system is conserved and is given by
\begin{equation}
   \mathcal{H} =  \sum_{i=1}^{n-1} \frac{p_i^2}{2} = E
\end{equation}
and assuming, for simplicity, that \ $E = 1/2$ \ energy unit, the
phase space of the system is reduced to \ $V^{n-1}\times S^{n-2}$.
\ We have tacitly assumed that there are no other integrals of
motion except the total energy \ $E$. \ As a result, the phase
space of the system becomes \ $V^{n-1}\times S^{n-2}$ \ but cannot
be further reduced due to the lack  of such another conserved
quantity. The part of phase space giving a non-trivial
contribution to the physical behavior of the system is \
$S^{n-2}$. \ This is equivalent to reducing the micro-canonical
distribution dependence on the momenta only \
\begin{equation}
 \tau(p_i, x_j) = \tau(p_i)
\end{equation}
Indeed, the factors due to \ $V^{n-1}$ \ are integration
constants, which can be safely
omitted in the subsequent arguments. \\

We now add one more particle to the gas of the \ $n-1$ \
particles. We assume that this particle which we add is stationary
in the frame or reference of the line segment. The total energy of
the system will remain \ $E$ \ and its phase space will now become
\ $V^n\times S^{n-1}$. \ Due to collisions with the other
particles, the one that we added will eventually acquire the same
average kinetic energy  \ $\overline{E}=\frac{E}{n}$ \ as the
other particles (equipartition). We consider the non-trivial part
of the micro-canonical expectation value
\begin{equation}
   \langle p_i^2 \rangle \ = \int_{S^{n-1}} p_i^2 \ \tau (p_j) \
   d\omega
\end{equation}
where \ $d\omega$ \ is the (normalized) Lebesgue measure on \
$S^{n-1}$ \ associated to the round metric. Since \ $S^{n-1}$ \ is
an isotropic space, we have
\begin{equation}
  \langle p_i^2 \rangle \ = \ \int_{S^{n-1}} \frac{1}{n} \sum_{i=1}^n p_i^2 \
  \tau (p_j) \ d\omega
\end{equation}
which, upon integration and after using (47) and (48), gives
\begin{equation}
   \langle p_i^2 \rangle \ = \ \frac{1}{n}
\end{equation}
The isotropy of \ $S^{n-1}$ \ also implies that \ $\langle p_i
\rangle = 0$, \ which gives for the standard deviation \
$\sigma_n$ \ of the momentum of each particle of the $n$-particle
gas
\begin{equation}
      \sigma_n \ = \ \frac{1}{\sqrt{n}}
\end{equation}
This result can be interpreted as follows: as the number of
particles \ $n$ \ of the gas increases, there is an uncertainty  \
$\sigma_n = n^{-\frac{1}{2}}$ \ in the value of the momentum of
each particle. If we add a stationary particle, the \ $n$-particle
gas will be indistinguishable from the \ $(n+1)$-particle gas
within an accuracy which is determined by \ $\sigma_n$. \ The
perturbation that the system will undergo by the introduction of
such a stationary particle will not be detectable within the given
level of accuracy. This is despite the fact that the phase spaces
\ $S^{n-1}$ \ and \ $S^n$ \ are not even homeomorphic, let alone
diffeomorphic. We can easily check that \ $S^n$ \ and \ $S^{n-1}$
\ are topologically, hence differentiably, distinct by comparing
their corresponding cohomology groups [10], for instance. The
relation of the metric properties of \ $S^n$ \ and \ $S^{n-1}$ \
can be found as follows: Let \ $ds_{n-1}, \ ds_n$ \ denote the
Riemannian round metrics of \ $S^{n-1}, \ S^n$ \ respectively.
Then topologically \ $S^n = \Sigma (S^{n-1})$ \ where \ $\Sigma$ \
in this relation indicates the spherical suspension [13]. Because
of this suspension, we come to suspect that \ $ds_n$ \ and \
$ds_{n-1}$ \ are related by
\begin{equation}
       ds_n \ = \ [0, \pi]\times_{\sin\theta} ds_{n-1}
\end{equation}
In (56), \ $\times_{\sin\theta}$ \ indicates a warped product with
function \ $\sin\theta$. \ We can rewrite (56), in a slightly more
familiar form as
\begin{equation}
       ds^2_n \ = \ (d\theta)^2 + \sin^2\theta  \ ds^2_{n-1}
\end{equation}
which can be straightforwardly seen to be correct. The case of a
particle  added to the system is expressed by a slowly varying
value of \ $\theta$ \ around \ $\theta=\frac{\pi}{2}$. \ Such an
approximation gives \ $d\theta = 0$ \ and up to first order in \
$\theta$ \ (which is proportional to \ $\sigma_n$) we find
\begin{equation}
     ds_n \ = \ ds_{n-1}
\end{equation}
which, in turn,  implies that the immersion \ $i:
S^{n-1}\hookrightarrow S^n$ \ is locally distance-preserving. \
Therefore, within a momentum uncertainty \ $\sigma_n$, \ it is
impossible to find locally any  difference between the metric
spaces \ $S^n$ \ and \ $S^{n-1}$. \ Roughly speaking, this amounts
to stating that if someone looks at \ $S^n$, \ $S^{n-1}$ \ with
eyeglasses whose resolution is larger than \ $\sigma_n =
n^{-\frac{1}{2}}$, \ then they appear to be
identical, as metric spaces, in a tubular neighborhood of \ $S^{n-1}$.\\

It is probably worth mentioning at this point that  similar
arguments can be presented [9] for systems whose configuration
space \ $\mathcal{M}_l$ \ is \ $l\in\mathbb{N}$ \ copies of \
$S^n$, \ namely
\begin{equation}
\mathcal{M}_l \ = \ \underbrace{S^n \times S^n \times \ldots
\times S^n}_{l \ \mathrm{times}}
\end{equation}
A physical example can be provided by \ $l$ \ different ideal
gases, each one containing $n$-particles which coexist in volume \
$V$. \ In such a case the phase space volume is the product of the
Riemannian volumes on each copy of \ $S^n$. \ Let \ $x =
(x_1,x_2,\ldots, x_l)$ \ and \ $y = (y_1,y_2,\ldots,y_l)$ \ be two
elements of \ $\mathcal{M}_l$. \ The distance \
$\rho_{\mathcal{M}_l}(x,y)$ \ between \ $x,y \in \mathcal{M}_l$ \
is chosen to be
\begin{equation}
 \rho_{\mathcal{M}_l}(x,y) \ = \ \left( \sum_{i=1}^l \rho
 (x_i,y_i)^2 \right)^\frac{1}{2}
\end{equation}
as one would normally expect for the Cartesian product of metric
spaces. It turns out, not too surprisingly, that the constants in
the L\'{e}vy-Gromov inequality are the same as the ones entering
(41) and the interpretation of momentum localization is
analogous to that case. \\


                              \vspace{5mm}

\centerline{\sc 5. \ Discussion and conclusions}

                               \vspace{3mm}

By using analytic and geometric arguments, we have described the
behavior of a special case of an ideal gas, and determined the
metric and measure-theoretical behavior of its phase space in the
thermodynamic limit. As a by-product of this approach, we obtained
an upper bound for the rate of approach of this system to
equilibrium. Although these analytic results have been known for a
while [3], we have not been able to find a reference that makes
explicit the approach, connections and
statistical implications that we make in the present letter. \\

The measure concentration that we observe in the case at hand
constitutes, in a way, a form of symmetry breaking. Indeed, we
start with \ $S^n$ \ (vacuum of the ``unbroken phase") equipped
with the round metric whose isometry group is \ $O(n)$. \ As \ $n$
\ increases, the Haar measure of \ $S^n$ \ localizes in a tubular
neighborhood of \ $S^{n-1}$ \ (vacuum of the ``broken phase"),
which has the smaller isometry group \ $O(n-1)$. \ In our case,
the parameter describing the concentration of measure, which plays
the role that the order parameter plays in the usual symmetry
breaking, is the dimension of the phase space \ $n$ \ of the
system. From the viewpoint of measure theory, a higher dimension,
which physically describes a larger number of degrees of freedom,
does not lead to a more complicated behavior, a fact which is
well-known and extensively employed when constructing the
thermodynamic limit of a statistical system. This behavior is also
evident in the simplifications occurring during the \ $1/N$ \
expansion of the \ $O(N)$ \ models [14] and the associated gauge
theories with a large number \ $N_c$ \ of colors [15], [16].
Needless to say, the thermodynamic limit of a statistical system
also leads to such simplifications. Actually, part of the goal of
this paper was the description, in a more geometric language, of
the way some systems, like our specifically chosen example of an
ideal gas, behave when considering their thermodynamic limit.\\

For the future, we would like to know whether the present analysis
can be extended to cover interacting statistical systems. This
would certainly be of much more interest than the example we have
presented in this paper. The difficulty with interacting systems
is threefold: First, they have phase spaces that are considerably
more complicated than those of free systems. Second, checking the
ergodicity of the Hamiltonian flow on such phase spaces, which
provides the dynamical foundation of statistical mechanics and its
mixing properties which determine the rate of approach of the
system toward equilibrium is practically impossible for all but
the simplest cases. Third, explicit calculation of the canonical
partition function of most interacting systems is practically
intractable, generically, except through perturbation theory or
lattice approximations. Even if such calculations were possible
however, it is unlikely that the final result could be expressed
in terms of purely, or easily identifiable, geometric quantities
as in the example we have analyzed. Then, instead of the
relatively straightforward analytic and geometric arguments
presented in this paper, the full machinery of measure theory
would have to be used, a fact that would enormously complicate the
analysis and detract from the underlying geometric structures. In
the spirit of the present paper, considerable progress in
generalizing the arguments and applying them to the case of spin
glasses was made in [17].\\

                               \vspace{5mm}


\noindent{\sc Acknowledgement:} \ We are grateful to the referee
whose criticisms helped correct some mistakes of an earlier
version of the manuscript and whose  comments helped
considerably elucidate several points of the exposition.\\

                                 \newpage

\centerline{\sc References}

                               \vspace{3mm}

\noindent [1] M. Ledoux, \ \emph{The Concentration of Measure
                     Phenomenon} \ AMS, Providence (2001)\\
\noindent [2] V.D. Milman, \ Funct. Anal. Appl. {\bf 5}, 28 (1971)\\
\noindent [3] P. L\'{e}vy, \ \emph{Problemes Concretes d'Analyse
                 Fonctionelle}, Gauthier-Villars, Paris (1951)\\
\noindent [4] M. Gromov, \ \emph{Metric Structures for Riemannian
              and Non-Riemannian Spaces}, \\
                \hspace*{5mm}  Birkh\"{a}user, \  Boston (1999)\\
\noindent [5] Yu.D. Burago, V.A. Zalgaller, \ \emph{Geometric
                  Inequalities}, \ Springer, Berlin (1988)\\
\noindent [6] I. Chavel, \ \emph{Isoperimetric Inequalities}, \
              Camb. Univ. Press, Cambridge (2001)\\
\noindent [7] M. Gromov, \emph{Isoperimetric Inequalities in
                            Riemannian Manifolds}, \  Appendix I of Ref.[9]\\
\noindent [8] T. Figiel, J. Lindenstrauss, V. Milman, \ Acta Math. {\bf 139}, 53 (1977)\\
\noindent [9] V.D. Milman, G. Schechtman, \ \emph{Asymptotic
                Theory of Finite Dimensional Normed \\
                   \hspace*{5mm} Spaces}, \ Lecture Notes in
                       Mathematics \ {\bf 1200}, \ Springer, \ Berlin (1986)\\
\noindent [10] V.D. Milman, \ Ast\'{e}risque, \ {\bf 157-8}, \  273 \ (1988)\\
\noindent [11] N. Kalogeropoulos, \ Int. J. Mod. Phys. A {\bf 23}, \ 509 \ (2008)\\
\noindent [12] I.S. Gradshteyn, I.M. Ryzhik, \ \emph{Table of
                         Integrals, Series and Products}, 5th Edition, \\
                               \hspace*{5mm} Academic Press, \ London (1984)\\
\noindent [13] E. H. Spanier, \ \emph{Algebraic Topology}, \ McGraw Hill, \  New York (1966)\\
\noindent [14] G. 't Hooft, \ \emph{Large N}, \ {\sf arXiv:hep-th/0204069}\\
\noindent [15] G. 't Hooft, \ \emph{Confinement at Large $N_c$},
                  \ Presented at \ ``Large N QCD", \\
                     \hspace*{6.5mm} Trento, Italy, 5-9 July 2004, \ {\sf
                        arXiv:hep-th/0408183}\\
\noindent [16] M. Moshe, J. Zinn-Justin, \ Phys. Rep. {\bf 385},
            69 \ (2003)\\
\noindent [17] M. Talagrand, \ IHES Publications, \ {\bf 81}, \ 73  \ (1995) \\

                               \vfill

\end{document}